# An Open Source P2P Encrypted Voip Application

Ajay Kulkarni
Operations & Cross Product Technology
Barclays Investment Bank
Pune, India

Saurabh Kulkarni
Software Engineer
Accenture
Mumbai, India

*Abstract*—Open source is the future of technology. This community is growing by the day; developing and improving existing frameworks and software for free. Open source replacements are coming up for almost all proprietary software nowadays. This paper proposes an open source application which could replace Skype, a popular VoIP soft phone. The performance features of the developed software is analyzed and compared with Skype so that we can conclude that it can be an efficient replacement. This application is developed in pure Java using various APIs and package and boasts features like voice calling, chatting, file sharing etc. The target audience for this software will initially only be organizations (for internal communication) and later will be released on a larger scale.

## I. INTRODUCTION

Email was the original killer application for the Internet. Today, voice over IP (VoIP) and instant messaging (IM) are fast supplementing email in both enterprise and home networks. Skype is an application that provides these VoIP and IM services in an easy to use package that works behind Network Address Translators (NAT) and firewalls. It has attracted a user-base of 70 million users, and is considered valuable enough that Microsoft recently acquired it for $8.5 billion [1]. In this paper, we present an open source replacement for Skype and a measurement study between the developed Peer-to-Peer application and Skype. While measurement studies of both P2P file sharing networks [2] and traditional VoIP systems [3] have been performed in the past, little is known about VoIP systems that are built using a P2Parchitecture.

One of our key goals in this paper is to understand how efficient this P2P VoIP application is to replace a giant like Skype which is also a P2P application. A peer-to-peer VoIP network typically consists of a core proxy network and a set of clients that connect to the edge of this proxy network (Fig. 6). This network allows a client to dynamically connect to any proxy in the network and to place voice calls to other clients on the network. VoIP uses the two main protocols: route setup protocol for call setup and termination, and Real-time Transport Protocol (RTP) [9] for media delivery. In order to satisfy QoS requirements, a common solution used in peer-to-peer VoIP networks is to use a route setup protocol that sets up the shortest route on the VoIP network from a caller source to a receiver destination. RTP is used to carry voice traffic between the caller and the receiver along an established bi-directional voice circuit.

Now, talking about the software license, as the developed software is open source its source code is freely available to all for further development. Its free availability gives scope for peer review, regular bug fixes and hence there is an increase in reliability of the application. Security flaws can be analyzed by anyone and can be fixed as and when a loophole is discovered. These are just some of the key points on why open source is preferred over proprietary software these days, the full list is endless.

Overall, this paper makes three contributions. First, light is shed on the VoIP network construction. Second, the architecture and design of the developed software is described in detail. Third, a comparison is done between the developed P2P VoIP application and Skype.

## II. RELATED WORK

Skype offers three services: VoIP allows two Skype users to establish two-way audio streams with each other and supports conferences, IM allows two or more Skype users to exchange small text messages in real-time, and file-transfer allows a Skype user to send a file to another Skype user (if the recipient agrees). Skype also offers paid services that allow Skype users to initiate and receive calls via regular telephone numbers through VoIP-PSTN gateways.

Despite its popularity, little is known about Skype's encrypted protocols and proprietary network. Skype is related to KaZaA; both the companies were founded by the same individuals, there is an overlap of technical staff, and that much of the technology in Skype was originally developed for KaZaA. Network packet level analysis of KaZaA [14] and of Skype [15] support this claim by uncovering striking similarities in their connection setup, and their use of a "supernode"-based hierarchical peer-to-peer network.

Supernode-based peer-to-peer networks organize participants into two layers: supernodes, and ordinary nodes. Such networks have been the subject of recent research in [16]. Typically, supernodes maintain an overlay network among themselves, while ordinary nodes pick one (or a small number of) supernodes to associate with.

## III. VOIP OVERVIEW

The section below describes the working of VoIP networks based on the function of the network components listed in Figure 7. Depending upon the particular network architecture [4] some of these network components [6] may be combined into a single solution.

*A. Call Agent/Sip Server/Sip Client*

The Call Agent/SIP Server/SIP Client is located in the service provider's network and provides call logic and call control functions, typically maintaining call state for every call in the network. The Call Agent will participate in signaling and device control, terminating or forwarding messages. There are numerous relevant protocols depending upon the network architecture including SIP (Session Initiation Protocol), SIP-T, H.323, BICC, H.248, MGCP/NCS, SS7, AIN, ISDN, etc. [19, 21]. A SIP Server provides equivalent function to a Call Agent in a SIP signaling network, its primary roles are to route and forward SIP requests, enforce policy (for example call admission control) and maintain call details records. For example the SIP Server in Service Provider 1's network will route and forward SIP requests from SIP Phones belonging to customers. A SIP Client provides similar function to a SIP Server, but originates or terminates SIP signaling rather than forwarding it to a SIP Phone or other Customer Premises Equipment. The Call Agent/SIP Server terminates the SIP signaling and converts it to H.248 or MGCP to set up a call to the correct subscriber. Call Agents are also known as Media Gateway Controllers, Soft switches and Call Controllers. All these terms convey a slightly different emphasis but maintaining call state is the common function reused with other services and to create new value added services.

*B. Service Broker*

The service broker is located on the edge of the service provider's service network and provides the service distribution, coordination, and control between application servers, media servers, call agents, and services that may exist on alternate technologies (i.e. Parlay Gateways and SCP's). The service broker allows a consistent repeatable approach for controlling applications in conjunction with their service data and media resources to enable services, to allow services to be reused with other services and to create new value added services.

*C. Application Server*

The Application Server is located in the service provider's network and provides the service logic and execution. Typically the Call Agent will route calls to the appropriate application server when a service is invoked that the Call Agent cannot support itself.

*D. Media Server*

This Media Server is located in the service provider's network. It is also referred to as an announcement server. For voice services, it uses a control protocol, such as H.248 or MGCP, under the control of the call agent or application server. Some of the functions the Media Server can provide are codec transcoding and voice activity detection, tone detection and generation and interactive voice response (IVR) processing.

*E. Signalling Gateway*

The Signaling Gateway is located in the service provider's network and acts as a gateway between the call agent signaling and the SS7-based PSTN. It can also be used as a signaling gateway between different packet-based carrier domains. It provides signaling translation, for e.g. between SIP and SS7 (Signaling System 7) or simply signaling transport conversion e.g. SS7 over IP to SS7 over TDM.

*F. Trunking Gateway*

The Trunking Gateway is located in the service provider's network and as a gateway between the carrier IP network and the TDM (Time Division Multiplexing)-based PSTN. It provides transcoding from the packet-based voice, VoIP onto a TDM network. Typically, it is under the control of the Call Agent / Media Gateway Controller (MGC) through a device control protocol such as H.248 or MGCP.

*G. Access Gateway*

The Access Gateway is located in the service provider's network. It provides support for POTS phones and typically, it is under the control of the Call Agent / Media Gateway Controller through a device control protocol such as H.248 or MGCP.

*H. Bandwidth Manager*

The Bandwidth Manager is located in the service provider's network and is responsible for providing the required QoS from the network. It is responsible for the setting up and tearing down of bandwidth within the network and for controlling the access of individual calls to this bandwidth.

*I. Bridge/Router*

The Bridge/Router is located at the customer premises and terminates the WAN (Wide Area Network) link at the customer premises. Voice services for example SIP phones, can be bridged/routed via this device.

*J. IP Phone/Microphone*

IP Phones and Microphones are located at customer premises and provide voice services. They interact with the Call Agent/SIP Server using a signaling protocol such as SIP, H.323 or a device control protocol such as H.248 or MGCP.

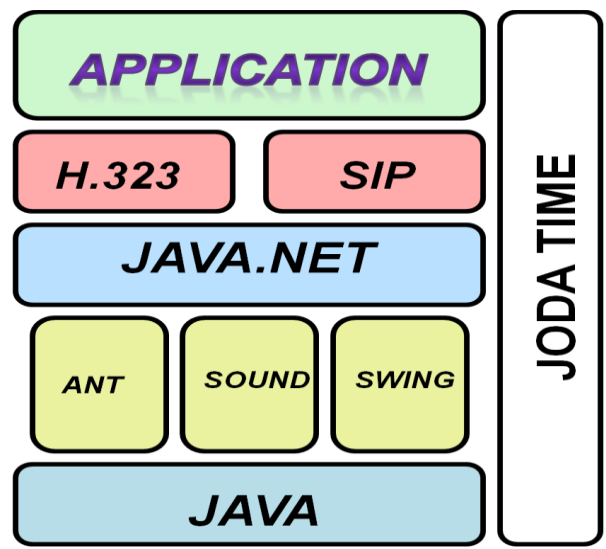

Fig. 1. Architecture of the developed software

## IV. THE VOIP OPEN SOURCE PROJECT

The open source software "movement" has received enormous attention in the last several years [11]. It is often characterized as a fundamentally new way to develop software [7] that poses a serious challenge [8] to the commercial software businesses that dominate most software markets today. The challenge is not the sort posed by a new competitor that operates according to the same rules but threatens to do it faster, better, cheaper. The developed open source application is developed as a replacement for proprietary VoIP software. The network architecture (Fig. 5) for the designed software is really simple to implement. Various Java APIs and packages like Swing, Java Sound API, Java.net package, Joda-Time API are used for the implementation of this project (Fig. 1). The Video Call feature is under development and will be released in version 2.0 of the software; features already implemented include P2P chatting, file sharing & encrypted voice call. Let us understand the implementation and function of every feature in the developed open source application. An extra feature of server monitoring is added for the organization centric release of this open source software.

### A. Database

The MySQL database is used to store all user data. The password is encrypted and stored along with the username (primary key); the profile picture of the user is stored in the form of BLOB (Binary Large Object) in the database.

### B. Login/Sign Up:

The login page in Fig. 2 is a simple form with two fields for username and password; when sign in is clicked a query runs on the database to check the validity of the credentials. The sign up page takes all necessary information from the user and creates an account for the user by storing his data securely in the database

### C. Home

All user functions are displayed on the home page in Fig. 3. The user can update his profile picture, check online users, start a chat session, and make a voice call and even share files with another user. To end the session, the user can click on sign out.

### D. Chat Box

A user can chat with multiple users at a time. The chat is implemented in the Java.net package and the delay is message exchange is virtually zero. The chat window is the big white box in Fig. 3.

### E. Encrypted Voice Call

The Java Sound API is the backbone of this feature giving all the necessary support to it. A custom voice call package has also been developed to encrypt voice data packets (using a custom encryption algorithm) [10, 21], improve sound quality during the call and also to minimize delay and the echo effect.

### F. File Share

Users can exchange files by clicking on the button on the left bottom of the home screen in Fig. 3, of any format except .exe between each other. The file sharing module is implemented using Java.net using simple TCP/IP port programming. The application has been tested for file sizes up to 20 MB and delay observed is negligible. The file is stored on the server for just a fraction of a second to prevent server overload.

### G. Chat Monitoring

This is not a P2P feature but it is specially included for the organization release of the software. The chat monitoring window helps organizations to keep a track of conversations between its employees for compliance purposes.

### H. Hardware

For the initial testing of the software, the application and database server was a remote computer with basic configurations. The client is really light and is currently supported only on Windows. This just shows how light and efficient open source applications are and why they are getting more and more popular every day.

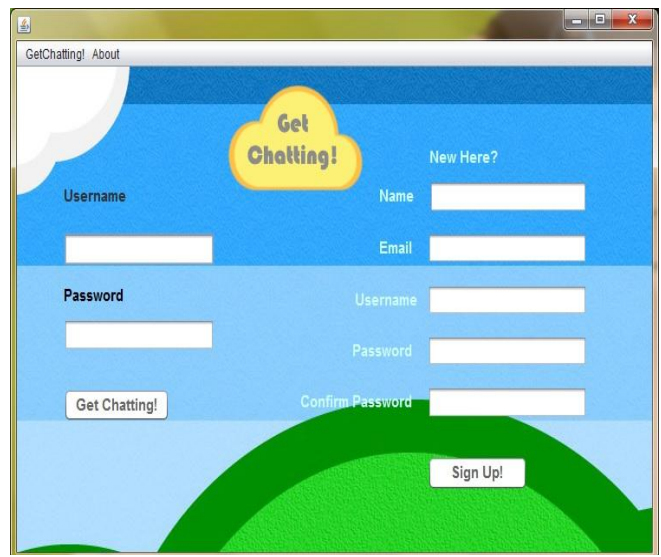

Fig. 2. Login screen of software

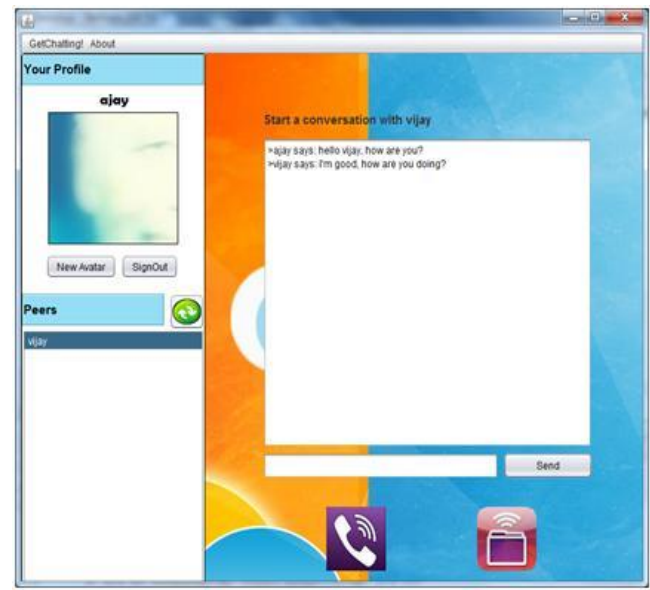

Fig. 3. Home screen of software

## V. OPEN SOURCE APPLICATION V/S SKYPE

Both the applications were tested on a network bandwidth of 2 Mbps and on a system with basic configurations. Experimental result [5, 12] for every feature is as follows:

### A. Voice Calling

The quality [18] of calling in both applications were analysed based on the clarity and delay in transmission of voice from source to destination and the results are presented in Table I.

TABLE I. VOICE QUALITY OF SKYPE VS. OS APPLICATION

| Application | Clarity | Delay |
|---|---|---|
| Skype | High | 22 ms |
| Open Source App | High | 31 ms |

### B. File Sharing

A standard file size of 2.5 Mb was used to test the results of file transfer. The upload and download speed is shown in Table II.

TABLE II. FILE SHARE SPEED OF SKYPE VS. OS APPLICATION

| Application | Upload Speed | Download Speed |
|---|---|---|
| Skype | 16 sec | 11 sec |
| Open Source | 19 sec | 13 sec |

### C. Chatting

P2P chatting is seamless in both applications with practically no delay. The organizational version of this feature in which the messages a routed through a monitoring server was also tested for delay and the results indicated that there is negligible delay as compared to the P2P version

### D. Handling Bulky Files – Stress Test

Both applications were stress tested by sending three files of size 25 MB each, back to back, after every 5 seconds. While the Skype window froze for a while but was back on track and started transmitting data, the Open Source Application crashed while sending the third file.

Thus stability is an issue which needs to be addressed in further releases.

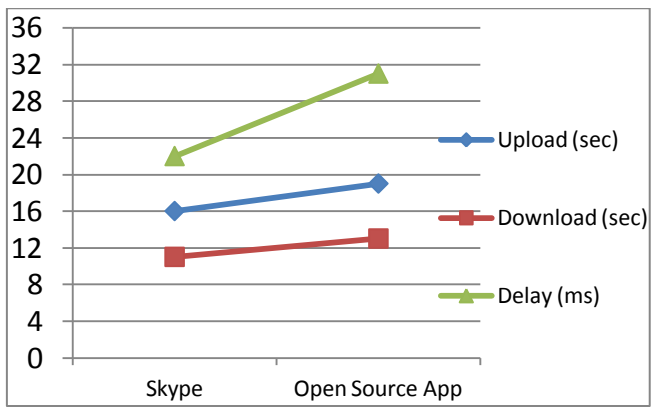

Fig. 4. Skype v/s Open source application performance

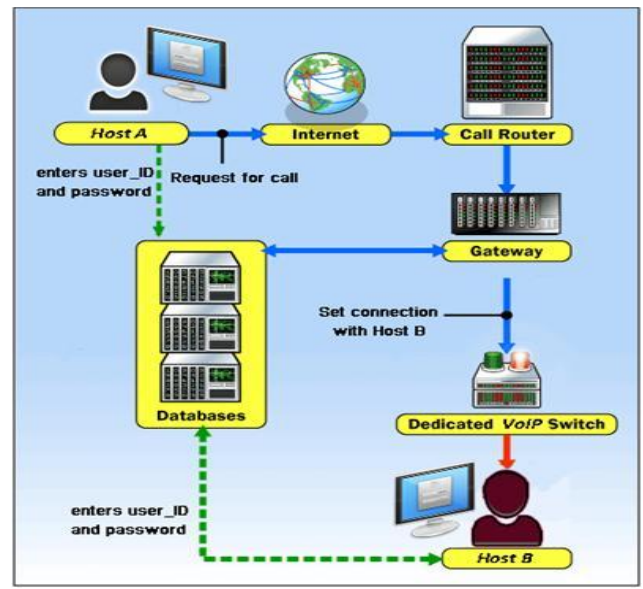

Fig. 5. System architecture of the open source application

## VI. FUTURE WORKS

The developed P2P open source application is currently being released for organizations and the full-fledged version for all users will be released in version 2.0. The Linux based version of the software is also in the works and will make its way out to user at the same time. Version 2.0 will also include features like Video Call [13] and Screen Share which will also be developed in Java. Users will also see an increase in call quality and numerous GUI tweaks. Also performance issues will be taken care of and random crashes due to large amounts of data transmission will be fixed in subsequent releases. The software will be licensed under the open source license and will be made public via a website which is also under development.

## VII. CONCLUSION

This paper presents an open source VoIP application to replace proprietary software like Skype. From the experimental data is gathered it is evident that the open source application can perform as good as Skype. The time delay in call routing and voice data transfer of the developed software is minimum considering that its voice data packets are being encrypted, and its performance can match that of Skype (refer Fig. 4). File sharing speeds are matching that of the proprietary software and hence the open source application performs up to the mark. The present experimental results are just preliminary in nature and further study is required on this topic. The GUI of Skype is really user friendly and has matured over the years; this is one area where the developed software has to catch up to a considerable extent.

Overall, the measurement data presented is useful for designing and modeling a peer-to-peer VoIP system. The architecture of the open source application in Fig. 5 lays down a foundation for all future VoIP system designing activities. The open source application can be further developed and tweaked by the community and can hopefully one day replace proprietary VoIP software.

REFERENCES

[1] Andrew Sorkin and Steve Lohr, Microsoft to Buy Skype for $8.5 Billion. *The New York Times* (May 10, 2011).

[2] Pouwelse, J., Garbacki, P., Epema, D., and Sips, H, The bittorrent p2p file-sharing system: Measurements and analysis. In *Proceedings of the IPTPS '05* (Ithaca, NY, Feb. 2005).

[3] Calyam, P., Sridharan, M., Mandrawa, W., and Schopis, P. Performance measurement and analysis of h.323 traffic. In *Proceedings of the 5th International Workshop on Passive and Active Network Measurement (PAM 2004)* (Antibes Juan-les-Pins, France, Apr. 2004).

[4] Cisco System, Data Considerations and Evolution of Transmission Network Design, http://www.cisco.com/en/US/prod/collateral/optical/ps5724/ps2006/prod_white_paper0900aecd803faf8f_ps2001_Products_White_Paper.html, 2009.

[5] The Network Simulator—ns2, http://www.isi.edu/nsnam/ns/, 2007.

[6] Paul Drew, Chris Gallon, Next-Generation VoIP Network Architecture, *MSF Technical Report*, March 2003.

[7] C. DiBona, S. Ockman, and M. Stone, Open Sources: Voices from the Open Source Revolution. Sebastopol, CA: O'Reilly, 1999. R. Nicole, "Title of paper with only first word capitalized," J. Name Stand. Abbrev.

[8] P. Vixie, "Software Engineering," in Open Sources: Voices from the Open Source Revolution, C. DiBona, S. Ockman, and M. Stone, Eds. Sebastopol, CA: O'Reilly, 1999, pp. 91-100. M. Young, The Technical Writer's Handbook. Mill Valley, CA: University Science, 1989.

[9] W. Mazurczyk and Z. Kotulski, "New Security and Control Protocol for VoIP Based on Steganography and Digital Watermarking," tech. rep., Institute of Fundamental Technological Research, Polish Academy of Sciences, June 2005, http://arxiv.org/ftp/cs/papers/0602/0602042.pdf

[10] D. Kundur and K. Ahsan, "Practical Internet Steganography: Data Hiding in IP," *Proc. Texas Wksp. Security of Information Systems*, Apr. 2003.

[11] Elliott M, Scacchi W. 2003. Free software developers as an occupational community: Resolving conflicts and fostering collaboration. In *Proceedings of the ACM International Conference on Supporting Group Work*, Sanibel Island, FL, November 2003, 21–30.

[12] Marc Greis, "Tutorial for Network Simulator NS", http://www.scribd.com/doc/13072517/tutorial-NS-full-byMARC-GREIS.

[13] Ajay Kulkarni, Saurabh Kulkarni, Ketki Haridas and Aniket More. Article: Proposed Video Encryption Algorithm v/s Other Existing Algorithms: A Comparative Study. International Journal of Computer Applications 65(1):1-5, March 2013. Published by Foundation of Computer Science, New York, USA

[14] Liang, J., Kumar, R., and Ross, K.W. The kazaa overlay: A measurement study. Computer Networks 49, 6 (Oct. 2005).

[15] Baset, S. A., and Schulzrinne, H. An Analysis of the Skype Peer-to Peer Internet Telephony Protocol. In Proceedings of the INFOCOM '06 (Barcelona, Spain, Apr. 2006).

[16] Xu, Z., and Hu, Y. Sbarc: a supernode based peer-to-peer file sharing system. In Proceedings of the 8th IEEE Symposium on Computers and Communications (ISCC'03) (Antalya, Turkey, July 2003).

[17] Castro, M., Costa, M., and Rowstron, A. Debunking some myths about structured and unstructured overlays. In *Proceedings of the NSDI '05* (Boston, MA, May 2005).

[18] S. Jadhav, H. Zhang, Z. Huang, ‖ Performance Evaluation of Quality of VoIP in WiMAX and UMTS‖ PDCAT (2011), pp. 378

[19] S. Sahabudin, M.Y. Alias. End-to-end delay performance analysis of various codecs on VoIP Quality of Service. Communications (MICC), 2009 IEEE 9th Malaysia International Conference on. vol., no., pp.607-612, 15-17 Dec. 2009.

[20] Yan Zhang and Huimin Huang, (2011) "VOIP voice network technology security strategies", Artificial Intelligence, Management Science and Electronic Commerce (AIMSEC), pp 3591-3594

[21] S. Ghosh, "Comparative Study of QOS Parameters of SIP Protocol in 802.11a and 802.11b Network", International journal of Mobile Network Communications & Telematics (IJMNCT), 2 (6), 21-30. (2012).

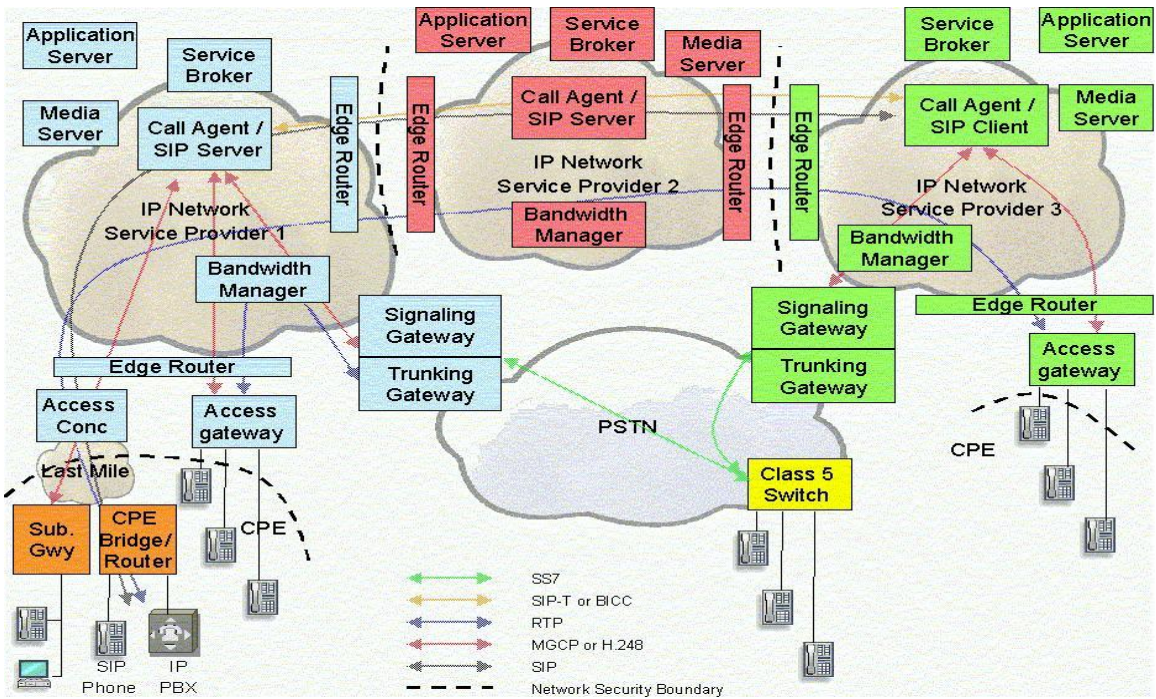

Fig. 6. Network architecture and components of a typical VoIP system